\documentclass[]{spie}  

\usepackage{amsmath,amsfonts,amssymb}
\usepackage{graphicx}
\usepackage[colorlinks=true, allcolors=blue]{hyperref}

\usepackage{setspace}
\usepackage{tocloft}
\usepackage[utf8]{inputenc}
\usepackage{rotating}
\usepackage{array}

\title{Centroiding and Extraction of Tip/Tilt Information from Nonlinear Curvature Wavefront Sensor Measurements}

\author[a]{Caleb G. Abbott}
\author[a]{Justin R. Crepp}
\author[b]{Stanimir O. Letchev}
\author[a]{Connor M. Smith}
\affil[a]{University of Notre Dame, Physics and Astronomy, Notre Dame, IN 46556, USA}
\affil[b]{Max-Planck Institute for Astronomy, Königstuhl 17, 69117 Heidelberg, Germany}
\authorinfo{Further author information: (Send correspondence to C.G.A.)\\C.G.A.: E-mail: cabbott3@nd.edu}

\begin{document} 
\maketitle

\begin{abstract}
The nonlinear curvature wavefront sensor (nlCWFS) uses multiple (typically four) out-of-focus images to reconstruct the phase and amplitude of a propagating light beam. Because these images are located between the pupil and focal planes, they contain tip/tilt information. Rather than using a separate sensor to measure image locations, it would be beneficial to extract tip/tilt information directly and routinely as part of the reconstruction process. In the presence of atmospheric turbulence, recovering precise centroid offsets for each out-of-focus image becomes a dynamic process as image structure is altered by changing aberrations. We examine several tip/tilt extraction methods and compare their precision and accuracy using numerical simulations. We find that the nlCWFS outer measurement planes confer more accurate and reliable tip/tilt information than the inner measurement planes, due to their larger geometric lever arm. However, in practice, finite field of view (detector region of interest) effects bias tip/tilt retrieval when using the outer planes due to diffraction. Using knowledge of the $z$-distance to each plane, we find that applying a best-fit linear model to multiple image centroid locations can offer fast and accurate tip/tilt mode retrieval. For the most demanding applications, a non-linear tip/tilt extraction method that self-consistently uses the speckle field may need to be developed. 
\end{abstract}



\section{INTRODUCTION}
\label{sec:intro}

The nonlinear Curvature wavefront sensor (nlCWFS) is an inherently sensitive device that offers a large capture range and ability to operate in the presence of scintillation \cite{Guyon2010,Mateen2011,Crass2014b,Crepp2020}. Due to these attributes, the nlCWFS represents a viable option for a variety of advanced adaptive optics (AO) applications including, but not limited to: astronomy, space domain awareness, remote sensing, laser communications, and related areas \cite{Ahn2023}. 

\begin{sidewaysfigure}
    \centering
    \includegraphics[width=1\linewidth]{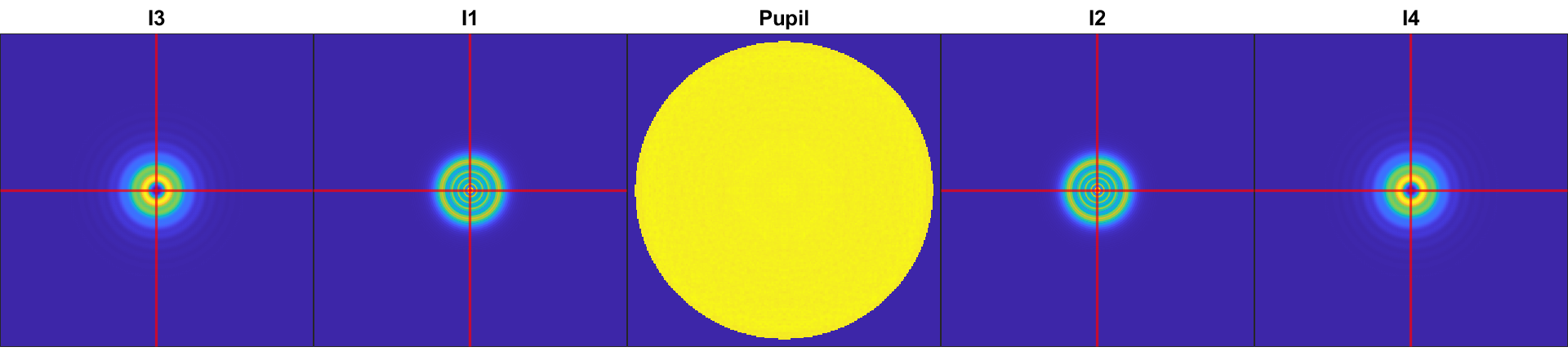}
    \includegraphics[width=1\linewidth]{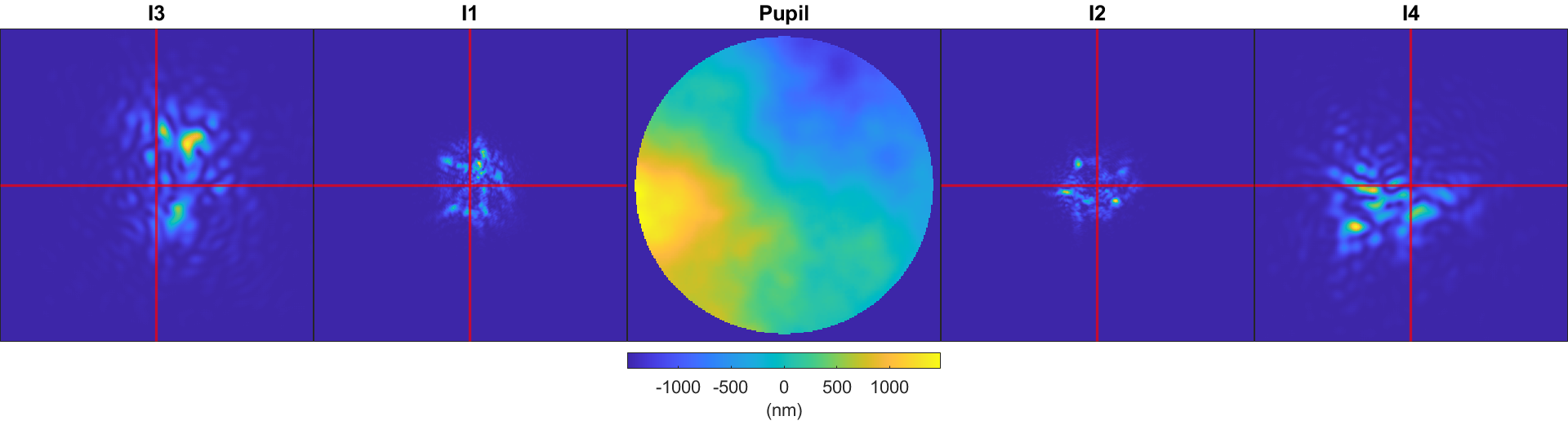}
    \caption{Diffracted light images measured at different $z$-distances from the pupil. (Top Row) Unaberrated beam. (Bottom Row) Aberrated beam with $D/r_0 = 8$. The center of each array (assuming no aberrations) is indicated by a red cross-hair. The location of the pattern depends on the amplitude of tip/tilt. This effect is seen in the bottom row whereby the aberrated images exhibit a downward trend from left to right due to the different path-lengths involved. Tip/tilt may be extracted using the relative shift from each measurement plane.}
    \label{fig:spots}
\end{sidewaysfigure}

Typically four measurement planes are used to provide sufficient path-length-diversity to reconstruct wavefront phase and amplitude information. In practice, accurate image centroiding plays an essential role for reliable wavefront reconstruction. Poor centroiding can negatively impact the reconstruction of higher-order modes by biasing the interpretation of diffracted light patterns, creating inconsistencies between planes. Tip/tilt modes cause a bulk horizontal/vertical translation of the centroid for each spot on the detector (Fig.~\ref{fig:spots}). 

By design, each measurement plane offers complementary information about the complex field (wavefront phase and amplitude). The two outer planes (further from the pupil, $I_3$ and $I_4$) are more sensitive to lower spatial frequency aberrations, while the two inner planes (closer to the pupil, $I_1$ and $I_2$) are more sensitive to higher spatial frequency aberrations. As such, the structure of each intensity pattern is generally expected to be asymmetric and unique (aberration-field dependent, path-length dependent, wavelength dependent, and time-dependent), making precise centroiding potentially challenging. This effect raises practical questions regarding the precision and accuracy of image centroiding techniques needed to effectively control tip/tilt. 

The goal of this project is to determine the amount of tip/tilt solely given the four nlCWFS intensity measurement planes. In this paper, we study several different methods for image centroiding including: brightest pixel, weighted average (center of light), and a best-fit procedure that combines multi-plane tip/tilt information content. Scenarios involving various levels of atmospheric turbulence strength are considered. Section \ref{sec:sims} describes the numerical simulations used to model nlCWFS measurements and methods for centroiding. Section \ref{sec:results} shows simulation results. Section \ref{sec:summary} provides a summary and concluding remarks.

\section{NUMERICAL SIMULATIONS AND METHODS}\label{sec:sims}

\subsection{Wavefront Sensing Model}\label{sec:model}
We model the nlCWFS using scalar wave optics simulations. The numerical methods used closely follow that of Letchev et al. 2023\cite{Letchev2023} and Potier et al. 2023\cite{Potier2023}, which we briefly summarize. A Kolmogorov spectrum of phase aberrations is generated using a single phase screen. The diameter of the telescope, $D$, and Fried parameter, $r_0$, are varied to study the response of the sensor as a function turbulence strength by changing the ratio $D/r_0$. The diameter of the telescope is set to be $D_{\rm tel}=1$ m and the Fried parameter is varied between $33 - 10$ cm. Varying $r_0$ allows us to vary the dynamic range of tip/tilt entering the system. The default pixel sampling used in simulations is a $1024 \times 1024$ grid. Appropriate zero-padding is used to minimize aliasing.


The model assumes monochromatic light at a wavelength of $\lambda=632$ nm. See Mateen et al (2011) and Letchev et al. (2022) for a discussion of polychromatic simulations and more\cite{Mateen2011,Letchev2022}. In this proceedings, we do not model wavefront compensation from a deformable mirror or fast steering mirror in closed loop; results for tip/tilt retrieval may thus be considered as a lower limit on performance. This topic will be covered in a follow-on study. 

A nlCWFS that uses four measurement planes located symmetrically along the $z$-axis from the optical system pupil is used to sense phase and amplitude of the aberrated wave. The $z$-distances of the planes are set using the optimal path-length criteria described in detail in Letchev et al. 2023 \cite{Letchev2023}. The planes located closest to the pupil are set to $z = \pm z_{\rm near}$, while the outer planes are at $z = \pm z_{\rm far}$. Light is propagated using physical optics programs in MATLAB using the Schmidt 2010 angular spectrum method to capture near-field diffraction effects \cite{Schmidt2010}. Wavefront reconstruction is performed using the Gerchberg-Saxton algorithm, modified to accommodate four intensity measurement planes \cite{Gerchberg1972,Guyon2010}. 

To study the precision and accuracy of tip/tilt sensing, we calculate centroids for each of the four diffracted light images. Unaberrated fields are used to calibrate the centroiding process; a flat incident wavefront would be seen as defocus at the different measurement planes, creating azimuthally symmetric images that can be centroided to arbitrary precision in the absence of noise (Top Row, Fig. ~\ref{fig:spots}). 

\subsection{Determining the True Tip/Tilt}\label{sec:tt_true}

\begin{figure}
    \centering
    \includegraphics[trim=2.4cm 5cm 0.6cm 3cm,clip=true,width=0.99\linewidth]{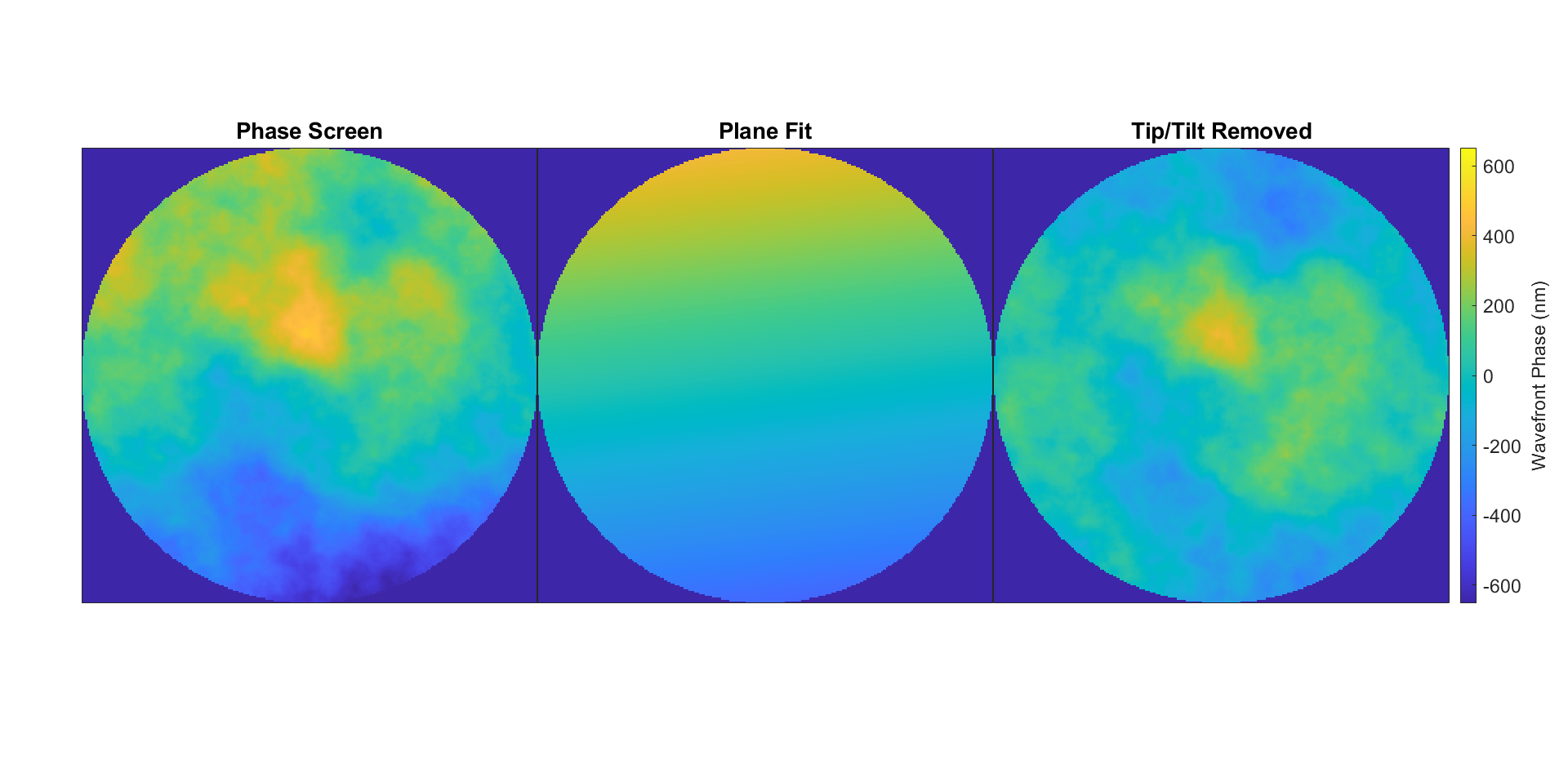}
    \caption{Example wavefront realization showing single phase screen generated using Kolmogorov statistics (left), the plane fit to the generated wavefront (middle), and the original phase screen with tip/tilt removed (right).}
    \label{fig:tt_calc}
\end{figure}

\begin{figure}
    \centering
    \includegraphics[width=0.78\linewidth]{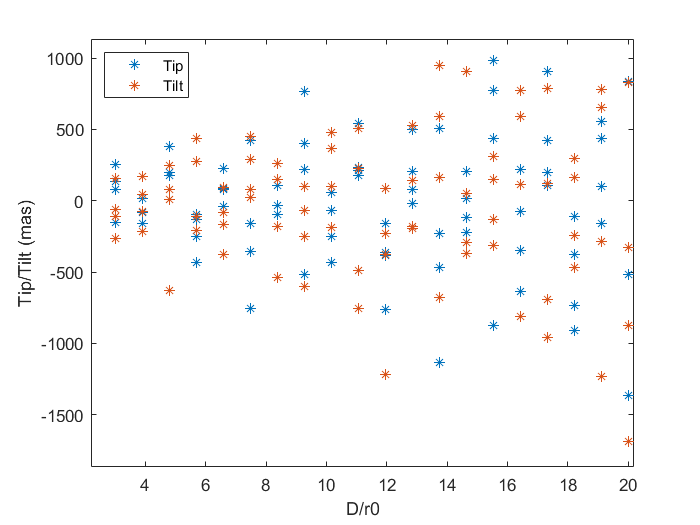}
    \caption{Tip and tilt truth values plotted as a function of simulated $D/r_0$. Such example realizations are used to quantify tip/tilt residuals using various centroiding methods. To help moderate memory usage, simulations in this paper span the range $3 \leq D/r_0 \leq 10$ unless otherwise noted.}
    \label{fig:tt_dr0}
\end{figure}

The true tip/tilt for each trial is determined by fitting a two-dimensional plane to the original wavefront. Plane fitting is performed by solving a least-squares problem that finds coefficients for a plane in three-dimensions,\cite{Schmidt2024}
\begin{equation}
\label{eqn:planefit}
z = ax + by + c.    
\end{equation} 
Figure~\ref{fig:tt_calc} shows an example of the procedure. Figure~\ref{fig:tt_dr0} shows the tip/tilt calculated for a range of $D/r_0$ values. 

After determining the true slopes for tip/tilt ($a$ and $b$ coefficients from Equation \ref{eqn:planefit}), we use the geometry of the sensor --- pixel scale ($\delta$), $z$-distance measured from the pupil, and an unaberrated beam for calibration --- to determine where diffracted light centroids should be located. Equations \ref{eqn:thetax} and \ref{eqn:xslopes} are used to calculate the central pixels of the aberrated images in each plane:
\begin{equation} \label{eqn:thetax}
    \theta_x = \arctan(a), \;\;\; \theta_y = \arctan(b)
\end{equation}
\begin{equation} \label{eqn:xslopes}
\begin{split}
    x_1 = x_0 - \frac{z_{\rm near} \tan(\theta_x)}{\delta}, \;\;\; y_1 = y_0 - \frac{z_{\rm near} \tan(\theta_y)}{\delta} \\
    x_2 = x_0 + \frac{z_{\rm near} \tan(\theta_x)}{\delta}, \;\;\; y_2 = y_0 + \frac{z_{\rm near} \tan(\theta_y)}{\delta}  \\
    x_3 = x_0 - \frac{z_{\rm far} \tan(\theta_x)}{\delta}, \;\;\; y_3 = y_0 - \frac{z_{\rm far} \tan(\theta_y)}{\delta} \\
    x_4 = x_0 + \frac{z_{\rm far} \tan(\theta_x)}{\delta}, \;\;\; y_4 = y_0 + \frac{z_{\rm far} \tan(\theta_y)}{\delta} 
\end{split}
\end{equation}
where $x_0$ and $y_0$ are the unaberrated central pixel locations, $z_{\rm near}$ and $z_{\rm far}$ are the distances of the planes from the pupil (near and far planes respectively), $\delta$ is the pixel scale in m/pixel, and $x_{1-4},y_{1-4}$ are the coordinates of the aberrated spot centers. These equations relate the image location for each wavefront to the true tip/tilt providing a basis from which to compare various centroiding methods.

\subsection{Defocus Plane Centroiding}\label{sec:centroiding}

\subsubsection{Brightest Pixel (BP)}\label{sec:bpix}
Centroiding using the Brightest Pixel (BP) found in a measurement plane is the most rudimentary but lowest latency method that we tested. BP simply identifies the pixel position with the greatest intensity and asserts the result as the image center. Given that large $D/r_0$ values greatly disrupt the shape of defocused-plane image patterns, often shifting the intensity of light away from geometric center, we might expect BP to offer less than ideal accuracy. Nevertheless, due to its fast and easy implementation, BP may prove beneficial as an initial guess for more sophisticated methods, so we tested its efficacy as a simplistic benchmark. 

\subsubsection{Weighted Average (WA)}\label{sec:com}

The next method that we tested was an intensity Weighted Average (WA). Although this approach does not incorporate the physics of diffraction, one would expect it to provide more consistent and accurate estimates than BP. The image centroid location, $x_n, y_n$, is estimated by
\begin{equation}
    x_n = \frac{\Sigma \; x_i I(x_i,y_j)}{\Sigma \; I(x_i,y_j)}, \;\;  y_n = \frac{\Sigma \; y_j I(x_i,y_j)}{\Sigma \; I(x_i,y_j)},
\end{equation}
for each plane ($n=1-4$) where the summation is performed over $i$, $j$ pixel indices for a given region of interest (ROI). Potential complications with the WA method (and BP method) include background contamination and intensity outliers. In the lab, ghost images caused by stray light or back-reflections can bias WA centroiding results without proper masking. Likewise, hot pixels or cosmic rays can influence WA values without proper filtering. In our simulations, we consider an ideal optical system and noiseless detector.

\subsubsection{Best-Fit Weighted Average (BFWA)}\label{sec:formod}

We define the Best-Fit Weighted Average (BFWA) as a WA centroiding method that uses the nlCWFS's geometry and layout to estimate the amount of tip/tilt in the system. First, the WA results for each pair of planes, $I_1$ with $I_2$ and $I_3$ with $I_4$, are fitted with a straight line. Then, tip and tilt are determined using the physical distances between planes. Fitting may also be performed globally across all four planes simultaneously. Note that in the absence of noise, fitting a slope between planes is analogous to averaging individual tip/tilt values calculated for each pair of planes. Therefore, we expect results for averaging WA pairs to be comparable to the BFWA method.

\section{RESULTS}\label{sec:results}

\subsection{Tip/Tilt from Individual Image Centroids}\label{sec:individual}
\begin{figure}[h!]
    \centering
    \includegraphics[trim=0.6cm 0cm 0.6cm 0.5cm,clip=true,width=0.49\linewidth]{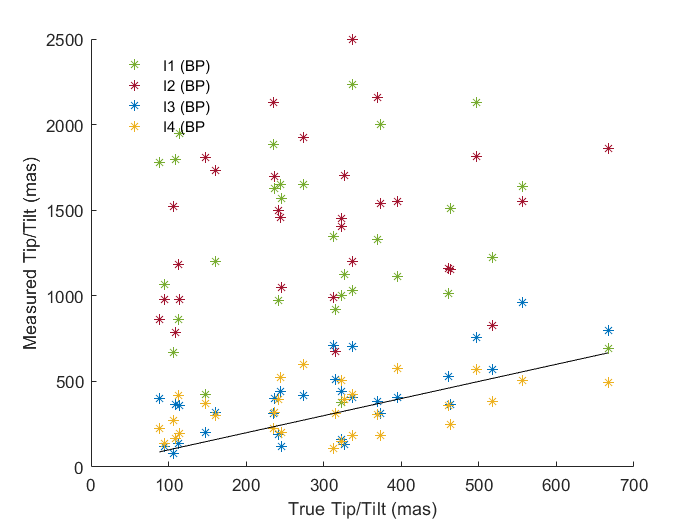}
    \includegraphics[trim=0.6cm 0.0cm 0.6cm 0.5cm,width=0.49\linewidth]{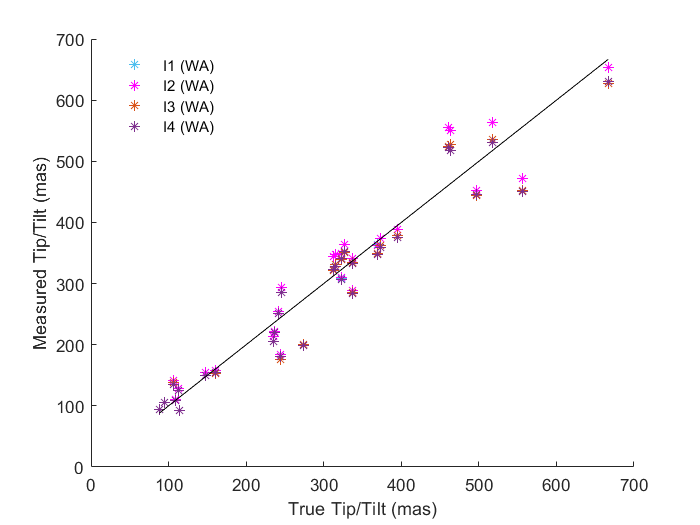}
    \caption{Measured tip/tilt (combined) versus true tip/tilt for the BP (left) and WA (right) methods.}
    \label{fig:bp_resi}
\end{figure}

Tip/tilt retrieval accuracy for the BP and WA methods using individual measurement planes are shown in Figure~\ref{fig:bp_resi}. Table \ref{tab:bpwa_resi} shows residual tip/tilt RMS values expressed in units of mas and $\lambda/D$. We find that the WA method is significantly more precise than tracking the BP. Typical centroiding scatter levels for the BP method are greater than the angular size of a diffraction-limited spot, whereas typical centroiding scatter levels for the WA method are smaller than the angular size of a diffraction-limited spot. 

The BP method may not allow for stable AO operation in practice. Depending on the application, the WA method may serve as a viable tip/tilt retrieval method. Results should improve when operating in closed-loop using a DM. Otherwise, more sophisticated approaches that combine measurement planes and/or take into consideration diffraction may need to be developed. 

\begin{table}[h!]
    \centering
    \begin{tabular}{c|cc|cc}
    \hline 
    \hline
    Image     & BP [mas] & WA [mas] & BP [$\lambda/D$] & WA [$\lambda/D$] \\ 
    \hline
     $I_1$    & 1143 & 40.4 & 8.79 & 0.31 \\ 
     $I_2$    & 1218 & 40.4 & 9.37 & 0.31 \\ 
     $I_3$    & 187 & 38.0 &  1.44 & 0.29 \\ 
     $I_4$    & 159 & 37.1 &  1.22 & 0.29 \\ 
     \hline
     \hline
    \end{tabular}
    \caption{Residual tip/tilt for each defocus plane comparing the brightest pixel (BP) and weighted average (WA) methods.}
    \label{tab:bpwa_resi}
\end{table}

\subsection{Tip/Tilt from Combining Image Centroids}\label{sec:averaging}
\begin{figure}
    \centering
    \includegraphics[width=0.75\linewidth]{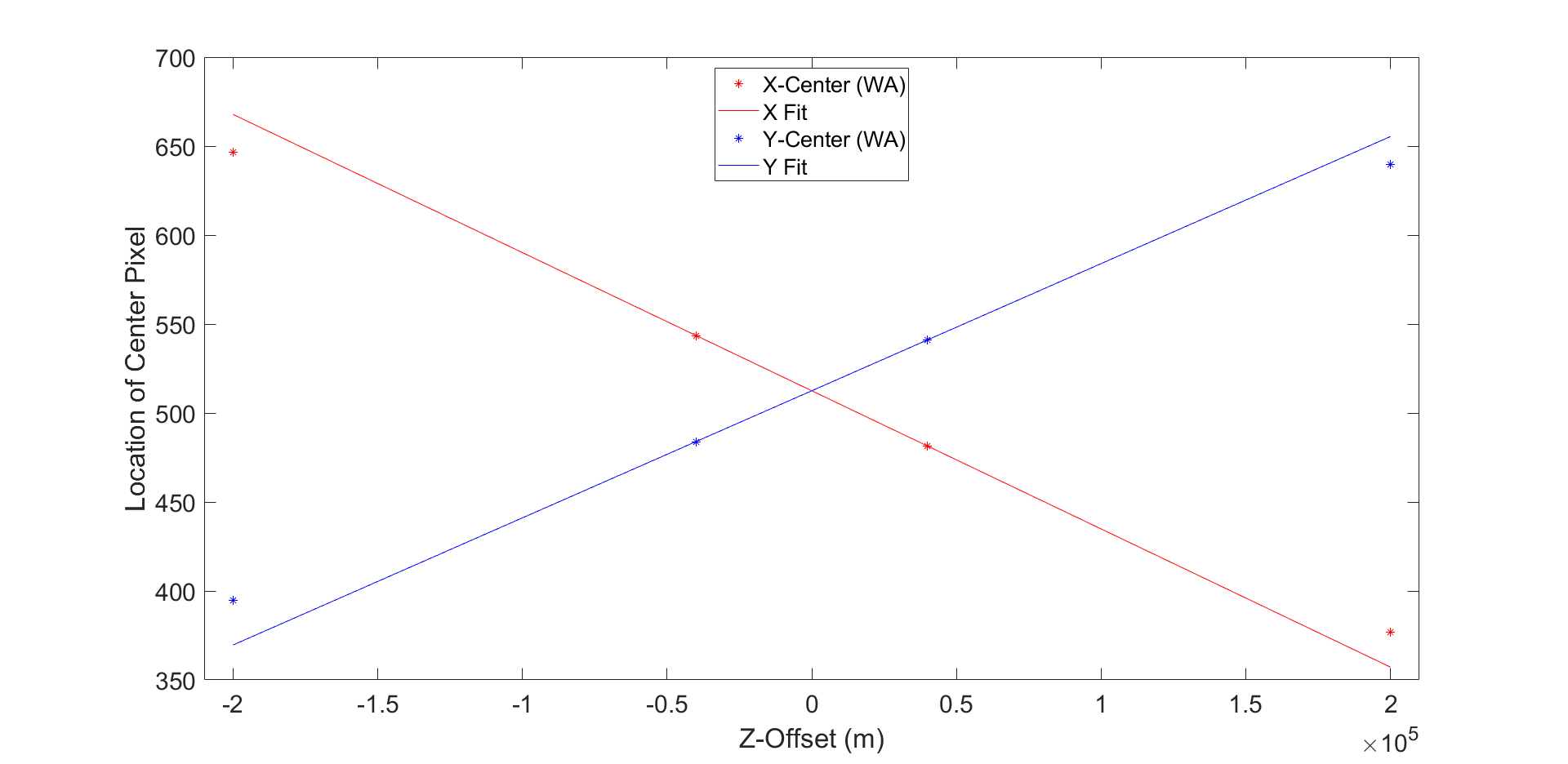}
    \includegraphics[width=0.75\linewidth]{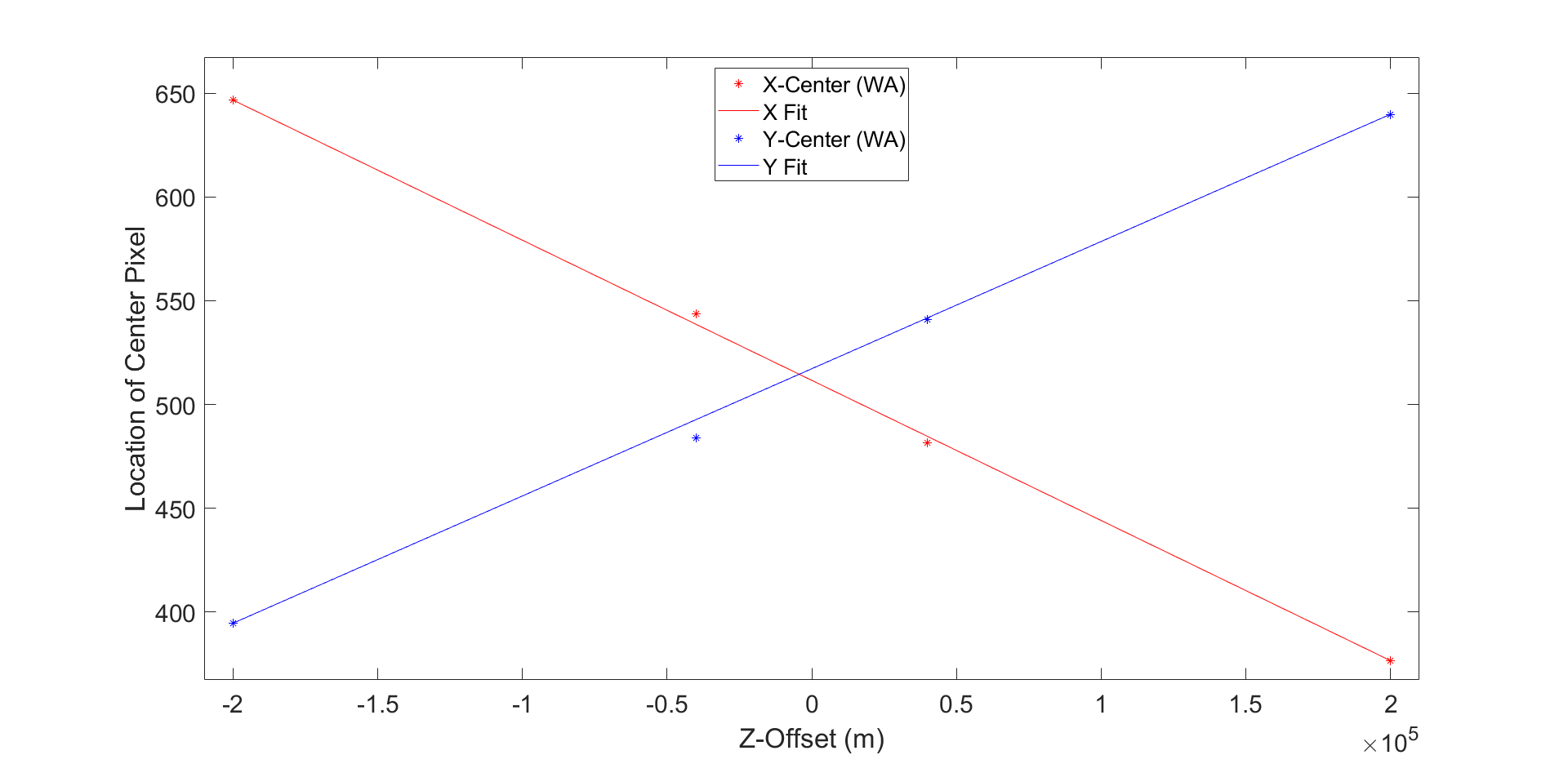}
    \includegraphics[width=0.75\linewidth]{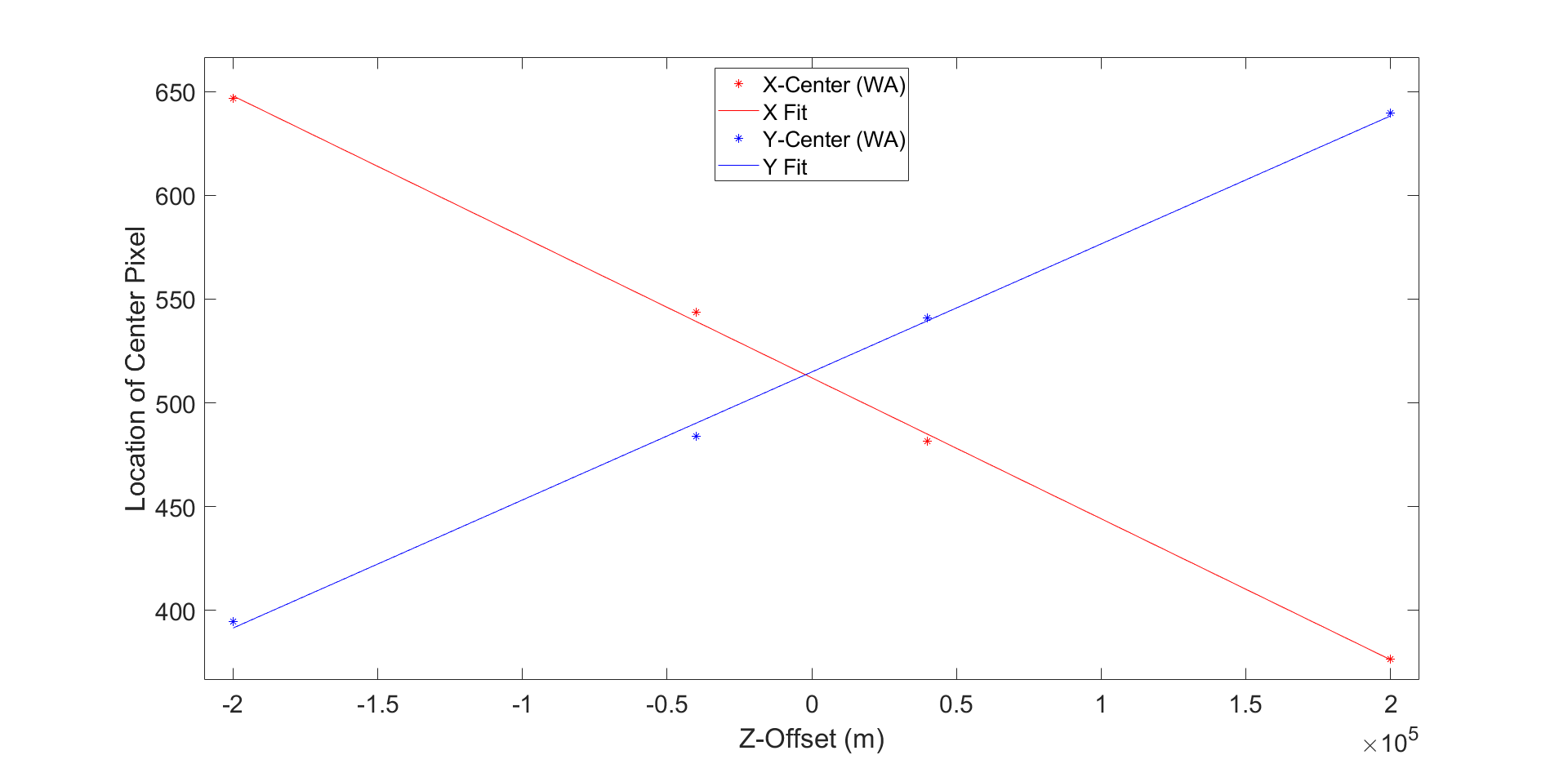}
    \caption{Linear fits to the WA centroids of the inner planes (top), outer planes (middle), and all planes (bottom). The red and blue points represent the central pixel ($x$ and $y$ axis respectively) as determined by the WA. The red and blue lines are best fit slopes to the various plane combinations (related to $\theta_x$ and $\theta_y$ respectively).}
    \label{fig:BFWA_slopes}
\end{figure}

Once the centroids from individual measurement planes have been estimated, there are a variety of options for deciding how to combine the results to estimate tip/tilt. We consider two approaches. The first approach averages tip/tilt estimates for pairs of planes (inner/outer) or all four planes simultaneously. The second approach is the best-fit WA (BFWA) described in $\S$\ref{sec:formod}. Figure~\ref{fig:BFWA_slopes} shows how information from different planes may be combined using the BFWA method. Table~\ref{tab:BFWA-table} summarizes results when combining image centroid measurements from different planes.

\begin{figure}[ht!]
    \centering
    \includegraphics[width=0.78\linewidth]{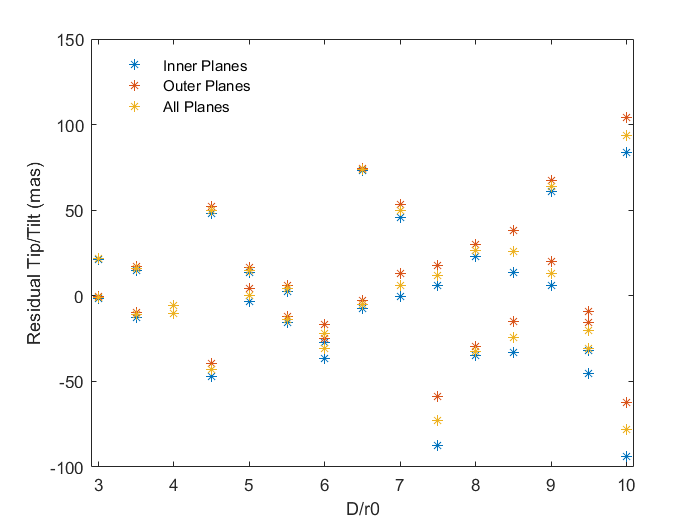}
    \caption{Tip/tilt residuals plotted against $D/r_0$ for the WA method.}
    \label{fig:resi_dr0}
\end{figure}

We find that the BFWA method offers minimal benefit over the WA method, with both techniques reaching a noise floor of $\delta \theta \approx 37$ mas. As alluded to earlier ($\S$\ref{sec:formod}), this result demonstrates that the two methods provide similar information in the absence of noise --- averaging is comparable to fitting a linear model when only utilizing two points. Figure \ref{fig:resi_dr0} shows the residuals for WA plotted against $D/r_0$ for each simulated wavefront. Insofar as diffraction creates non-uniform illumination patterns at the measurement planes, the scatter in tip/tilt is set by a combination of higher order wavefront errors, propagation distance, and pixel sampling. 




\begin{table}[ht!]
    \centering
    \begin{tabular}{c|ccc|ccc}
    \hline 
    \hline
    Plane Combination & BP [mas] & WA [mas] & BFWA [mas] & BP [$\lambda/D$] & WA [$\lambda/D$] & BFWA [$\lambda/D$] \\
    \hline
     Inner Planes &     1129     & 40.4   & 40.4  & 8.68 & 0.31 & 0.31 \\
     Outer Planes &     129      & 37.6   & 37.6  & 0.99 & 0.29 & 0.29 \\
     All Planes   &      604     & 38.4   & 37.6  & 4.65 & 0.30 & 0.29  \\
    \hline 
    \hline
     \end{tabular}
    \caption{Residual tip/tilt averaged over groups of planes comparing the brightest pixel (BP), weighted average (WA), and best-fit weighted average (BFWA) methods.}
    \label{tab:BFWA-table}
\end{table}

\begin{figure}[!h]
    \centering
    \includegraphics[width=0.99\linewidth]{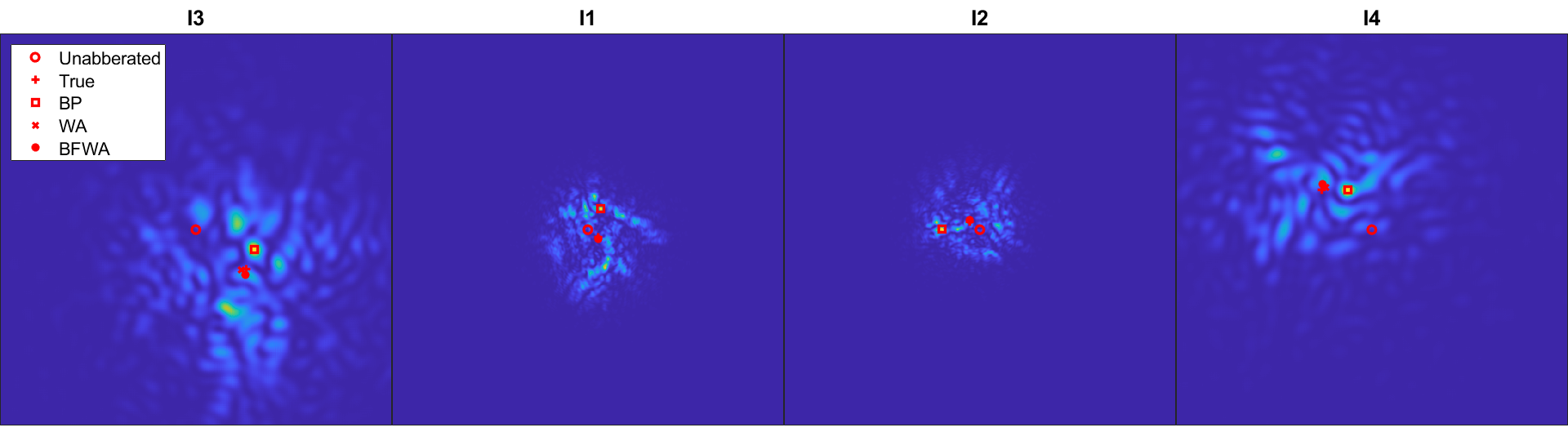}
    \includegraphics[width=0.99\linewidth]{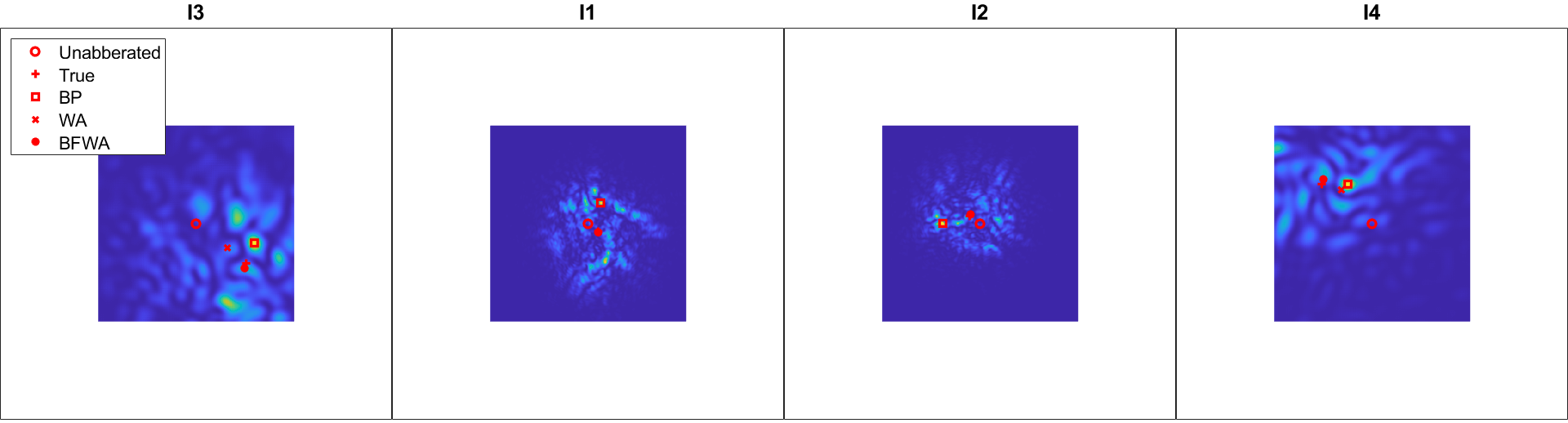}
    \caption{Images showing an unbiased detector (top row) and the effect of truncating the ROI (bottom row). The red points represent various centroids as determined by the methods in Section \ref{sec:centroiding}. The outer measurement planes ($I_3$ and $I_4$) show the greatest change as they are more-so affected by diffraction when limiting the ROI.}
    \label{fig:trunc_example}
\end{figure}

\subsection{Limited Region of Interest (ROI)} \label{sec:highdr0}
\begin{figure}
    \centering
    \includegraphics[width=0.69\linewidth]{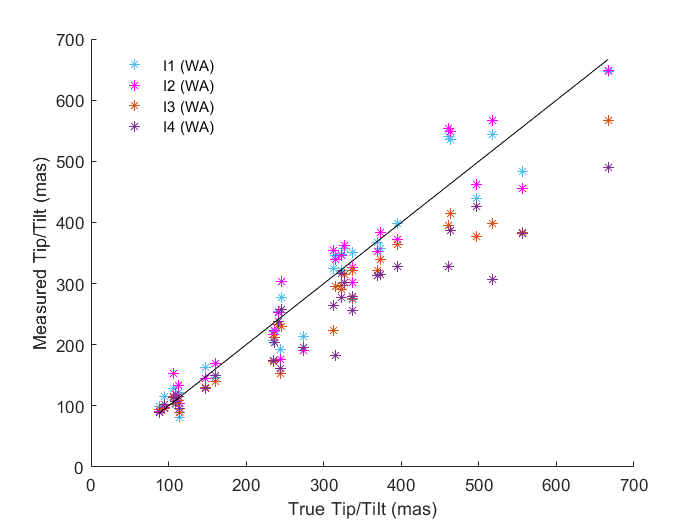}
    \caption{Tip/tilt bias caused by finite ROI. Results become systematically under-estimated when the outer-planes no longer fully collect diffracted light. \\}
    \label{fig:wa_res_lim}
\end{figure}

In practice, the nlCWFS must read out reasonably small ROI's from the detector to run the AO system at high frame rates. An example of truncated ROI's is shown in Figure \ref{fig:trunc_example}. To study the effects of limiting the ROI, we performed another set of simulations with a smaller array size ($512 \times 512$) using the same aberrations ($D/r_0 = 3-10$). Figure \ref{fig:wa_res_lim} shows the results of WA tip/tilt estimation for individual spots. 

As the tip/tilt angle increases, estimates for centroid locations of the outer planes become biased due to information loss. The results are systematically low due to the truncation of diffracted light. Thus, in the case of finite ROI, the most accurate estimate of tip/tilt is found by averaging the centroids of the inner planes via the WA or BFWA methods because they have a shorter propagation distance (Table \ref{tab:bpwa_resi_roi}, Table \ref{tab:BFWA-table-ROI}). This result is the opposite of previous examples that used an effectively unlimited ROI. 

\begin{table}[ht!]
    \centering
    \begin{tabular}{c|cc|cc}
    \hline 
    \hline
    Image     & BP [mas] & WA [mas] & BP [$\lambda/D$] & WA [$\lambda/D$] \\ 
    \hline
     $I_1$    & 1145 & 36.4 & 8.81 & 0.28 \\ 
     $I_2$    & 1217 & 42.9 & 9.36 & 0.33\\ 
     $I_3$    & 187 & 61.9  & 1.44 & 0.48\\ 
     $I_4$    & 159 & 81.2  & 1.22 & 0.62\\ 
     \hline
     \hline
    \end{tabular}
    \caption{Residual tip/tilt for each defocus plane comparing the brightest pixel (BP) and weighted average (WA) methods in a ROI limited scenario.}
    \label{tab:bpwa_resi_roi}
\end{table}

\vspace{0.4in}

\begin{table}[ht!]
    \centering
    \begin{tabular}{c|ccc|ccc}
    \hline 
    \hline
    Plane Combination & BP [mas] & WA [mas] & BFWA [mas] & BP [$\lambda/D$] & WA [$\lambda/D$] & BFWA [$\lambda/D$] \\
    \hline
     Inner Planes &     1130     & 38.6   & 38.6  & 8.69 & 0.30 & 0.30 \\
     Outer Planes &     129      & 69.5   & 69.5  & 0.99 & 0.53 & 0.53 \\
     All Planes   &      605     & 42.8   & 66.5  & 4.65 & 0.33 & 0.51 \\
    \hline 
    \hline
    \end{tabular}
    \caption{Residual tip/tilt averaged across multiple planes comparing the brightest pixel (BP), weighted average (WA), and best-fit weighted average (BFWA) methods in a ROI limited scenario.}
    \label{tab:BFWA-table-ROI}
\end{table}
\section{SUMMARY AND CONCLUDING REMARKS}\label{sec:summary}

We have explored several methods for accurately retrieving wavefront tip/tilt information from the nlCWFS. Using simulations, we show that tip/tilt may be recovered at a level comparable to the diffraction-limited spot size (tens of mas) of a telescope at visible or NIR wavelengths. Using a low-latency linear retrieval analysis, we find that practical limitations regarding ROI can bias tip/tilt estimates in the two outer measurement planes of the nlCWFS; this result is caused by diffraction when running the system without compensation from a deformable mirror. In such cases, it may be advisable to use the two inner measurement planes, despite their having a shorter geometric lever arm. 

Next steps in the analysis process involve modeling a closed-loop AO system while incorporating the effects of diffraction into tip/tilt retrieval. Self-consistently accounting for the speckle field in each plane is a non-linear process that should improve tip/tilt estimates at the expense of higher latency. In a follow-on article, we also aim to inject realistic sources of noise into the system to explore how photon statistics and detector read noise impact tip/tilt retrieval. Although a spatial domain error budget has been developed for the nlCWFS,\cite{Potier2023} the effects of photon noise and read noise have only been studied for reconstruction of higher-order modes. Experimental validation of these numerical predictions at the Notre Dame AO lab are on-going. 

\appendix   
\acknowledgments 
 
This research was supported in part by the Air Force Office of Scientific Research (AFOSR) grant number FA9550-22-1-0435. JC acknowledges support from the Naval Research Lab (NRL) summer faculty fellowship program. We acknowledge support from Northrop Grumman Space Systems.

\bibliography{abbott} 
\bibliographystyle{spiebib} 

\end{document}